\documentclass[12pt, letterpaper, preprint, comicneue]{aastex63}
\pdfoutput=1
\usepackage[T1]{fontenc}
\usepackage{color}
\usepackage{amsmath}
\usepackage{natbib}
\usepackage{ctable}
\usepackage{bm}
\usepackage[normalem]{ulem} 
\usepackage{xspace}
\usepackage{csvsimple} 

\usepackage{graphicx}
\usepackage{pgfkeys, pgfsys, pgfcalendar}

\let\oldbibliography\thebibliography 
\renewcommand{\thebibliography}[1]{%
  \oldbibliography{#1}%
  \setlength{\itemsep}{0pt}%
  \setlength{\parsep}{0pt}%
  \setlength{\parskip}{0pt}%
  \setlength{\bibsep}{0ex}
  \raggedright
}
\newcommand{\given}{\,|\,}

\newcommand{\bfi}[1]{\textbf{\textit{#1}}}

\newcommand{\eg}{\emph{e.g.}}
\newcommand{\ie}{\emph{i.e.}}

\newcommand{\bitem}{\begin{itemize}}
\newcommand{\eitem}{\end{itemize}}
\newcommand{\beq}{\begin{equation}}
\newcommand{\eeq}{\end{equation}}

\definecolor{orange}{rgb}{1,0.5,0}

\begin{document} \sloppy\sloppypar\frenchspacing 

\title{Cosmology with Galaxy Photometry Alone}

\newcounter{affilcounter}
\author[0000-0003-1197-0902]{ChangHoon Hahn}
\altaffiliation{Corresponding author}
\altaffiliation{changhoon.hahn@princeton.edu}
\affil{Department of Astrophysical Sciences, Princeton University, Peyton Hall, Princeton NJ 08544, USA} 

\author[0000-0002-4816-0455]{Francisco Villaescusa-Navarro}
\affil{Center for Computational Astrophysics, Flatiron Institute, 162 5th Avenue, New York, NY, 10010, USA}
\affil{Department of Astrophysical Sciences, Princeton University, Peyton Hall, Princeton NJ 08544, USA} 

\author[0000-0002-8873-5065]{Peter Melchior}
\affil{Department of Astrophysical Sciences, Princeton University, Peyton Hall, Princeton NJ 08544, USA} 
\affil{Center for Statistics \& Machine Learning, Princeton University, Princeton, NJ 08544, USA}

\author[0000-0001-7689-0933]{Romain Teyssier}
\affil{Department of Astrophysical Sciences, Princeton University, Peyton Hall, Princeton NJ 08544, USA} 
\affil{Program in Applied and Computational Mathematics,
Princeton University, Fine Hall Washington Road,       
Princeton NJ 08544-1000 USA}

\begin{abstract}
    We present the first cosmological constraints using only the observed
    photometry of galaxies. 
    \cite{villaescusa-navarro2022} recently 
    demonstrated that the internal physical properties of a single simulated galaxy 
    contain a significant amount of cosmological information.
    These physical properties, however, cannot be directly measured from
    observations. 
    In this work, we present how we can go beyond theoretical demonstrations to
    infer cosmological constraints from actual galaxy observables (\emph{e.g.}
    optical photometry) using neural density estimation and the CAMELS suite of 
    hydrodynamical simulations. 
    We find that the cosmological information in the photometry of a single
    galaxy is limited.
    However, we combine the constraining power of photometry from many
    galaxies using hierarchical population inference and place significant
    cosmological constraints.
    With the observed photometry of $\sim$20,000 NASA-Sloan Atlas galaxies, we 
    constrain $\Omega_m = 0.323^{+0.075}_{-0.095}$ and 
    $\sigma_8 = 0.799^{+0.088}_{-0.085}$. 
\end{abstract}


\section{Introduction} \label{sec:intro} 
In recent work, \cite{villaescusa-navarro2022}
showed that it is possible to place cosmological constraints by utilizing only the
internal properties of a single simulated galaxy.
They used galaxies from 2,000 state-of-the-art hydrodynamical simulations with
different cosmologies and astrophysical models from the
CAMELS project~\citep{villaescusa-navarro2021, villaescusa-navarro2022a} to train
moment networks~\citep{jeffrey2020a} that predict cosmological parameters from
galaxy properties. 
With only a handful of galaxy properties, including stellar mass ($M_*$),
stellar metallicity ($Z_*$), and maximum circular velocity ($V_{\rm max}$),
they were able to constrain $\Omega_m$ to 10\% precision with a single galaxy.
They found similar constraining power for galaxies simulated using the subgrid
physics models of the IllustrisTNG~\citep{pillepich2018, weinberger2018} and
SIMBA~\citep{dave2019} simulations. \cite{echeverri2023} reached similar 
conclusions using yet another subgrid physics model from 
Astrid~\citep{Astrid, Bird_2022}.

According to \cite{villaescusa-navarro2021}, the
cosmological information is derived from the imprint of $\Omega_m$ on the dark
matter content of galaxies that affects galaxy properties in a way distinct from
astrophysical processes.
Because $\Omega_b$ is fixed in CAMELS, which is justified by the tight
constraints from Big Bang 
Nucleosynthesis~\citep{aver2015, cooke2018, schoneberg2019}, galaxy properties
actually measure the baryon fraction, $\Omega_b/\Omega_m$\footnote{
In upcoming work, \cite{Cosmo1gal_SB28} find that galaxy properties can constrain 
$\Omega_m$ independently from $\Omega_b/\Omega_m$ using information on the age of 
the galaxy's stellar population and its star formation history.}.
For instance, $V_{\rm max}$ measures the depth of the total matter
gravitational potential while other properties like $M_*$ and $Z_*$ measures
the mass in baryons, so together they constrain the ratio
$\Omega_b/\Omega_m$.
In fact, a similar approach was already used by 
\cite{white1993} to constrain $\Omega_b/\Omega_m$  using galaxy
clusters. 

Despite promising signs that they may be useful cosmological probes, galaxy
properties themselves are {\em not} actually observable.
They are derived quantities that are typically inferred from photometry or
spectra and require modeling the spectral energy distribution
(SED) or emission lines~\citep[see][for a review]{conroy2013}.
In this work, we go beyond the theoretical considerations of previous works and
infer cosmological parameters from actual galaxy observables --- optical  
photometry.  
We leverage a simulation-based inference method that employs neural density 
estimation (NDE) to estimate the posterior of cosmological 
parameters given galaxy photometry, similar to the approach of \citet{hahn2022a}. 
Furthermore, since we expect a limited amount of cosmological information from
the photometry of a single galaxy, we present a hierarchical population
inference approach for inferring the posterior of cosmological parameters
from the photometry of multiple galaxies. 
Lastly, we present the cosmological constraints derived from applying this
approach to the photometry of $\sim$20,000 SDSS galaxies from the NASA-Sloan
Atlas. 

\section{Observations: NASA-Sloan Atlas}  \label{sec:nsa}
In this work, we analyze galaxy photometry from the NASA-Sloan
Atlas\footnote{\url{http://www.nsatlas.org/}} (hereafter NSA).
The NSA provides photometry of $z < 0.05$ galaxies observed by the Sloan
Digital Sky Survey~\citep[SDSS;][]{aihara2011} Data Release 8 with improved
background subtraction~\citep{blanton2011}. 
We use optical $g$, $r$, $i$, $z$ band absolute magnitudes derived using 
{\sc kcorrect}~\citep{blanton2007}, assuming a
\cite{chabrier2003} initial mass function (IMF). 

\begin{figure}[ht]
\begin{center}
    \centerline{\includegraphics[width=0.5\textwidth]{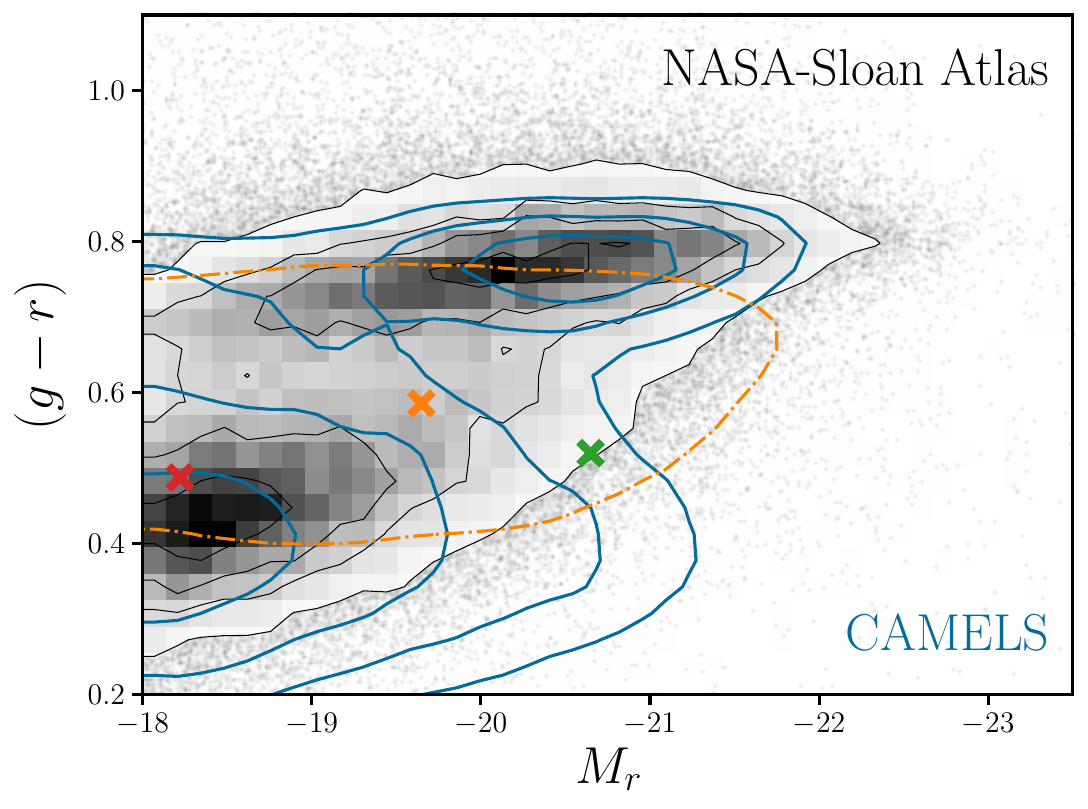}} 
    \caption{Color-magnitude distribution, $(g-r) - M_r$, of observed galaxies
    from the NSA (black) and simulated galaxies from CAMELS-TNG (blue). 
    Overall, the distributions of NSA and CAMELS-TNG galaxies are in good
    agreement. 
    The distribution for CAMELS-TNG is significantly broader because its
    galaxies are simulated using a wide range of cosmological and baryonic
    feedback parameters.
    We mark the 95$^{th}$ percentile contour of the NSA subsample of 
    22,338 galaxies analyzed in this work (orange dot dashed; Section~\ref{sec:nsa}) 
    along with three arbitrarily selected galaxies from this sample (crosses).
    }\label{fig:nsa}
\end{center}
\end{figure}

Out of the full NSA sample, we focus on luminous galaxies with 
$-18 > M_r > -22$ that have
precisely measured photometry: 
magnitude uncertainties below $(\sigma_g, \sigma_r, \sigma_i) < 0.022$ and 
$\sigma_z < 0.04$. 
In addition, we impose color cuts, specified in Table~\ref{tab:color}, to ensure 
that our observed sample is contained within the photometric distribution 
(\ie~support) of our simulated galaxies (Section~\ref{sec:sims}).
These cuts exclude NSA galaxies outside of the central 68 percentile range of the 
$g-r$, $g-i$, $g-z$, $r-i$, $r-z$, $i-z$ simulated galaxy color distributions.
The color cuts also remove NSA galaxies that potentially have observational
artifacts or problematic photometry.
We mark the 95$^{th}$ percentile contour of our NSA subsample in
Figure~\ref{fig:nsa} (orange dot-dashed).
In total, we use 22,338 NSA galaxies.

\section{Forward Model: CAMELS} \label{sec:sims}
We use simulated galaxies from CAMELS, a suite of hydrodynamical
simulations constructed over a wide range of cosmological and
hydrodynamical parameters.
In particular, we use the 1,000 hydrodynamical simulations constructed using
the state-of-the-art IllustrisTNG model (CAMELS-TNG). 
The simulations are generated with different cosmological parameters,
$\Omega = \{\Omega_m, \sigma_8\}$, and baryonic feedback parameters, 
$\mathcal{B} = \{A_{\rm SN1}, A_{\rm SN2}, A_{\rm AGN1}, A_{\rm AGN2}\}$,
arranged in a latin hypercube. 
$A_{\rm SN1}$ and $A_{\rm SN2}$ represent the normalization factors for the 
galactic wind flux and speed; 
$A_{\rm AGN1}$ and $A_{\rm AGN2}$ represent the normalization factors for the 
energy output and specific energy of AGN feedback.

In the 1,000 simulations, there are a total of $\sim$700,000 galaxies with 
$M_* > 2\times10^8 M_\odot/h$. 
The individual simulations, however, have different numbers of galaxies. 
For instance, simulations constructed at higher $\Omega_m$ values generally have
more galaxies.  
This parameter dependence means that the $\Omega$ and $\mathcal{B}$ distribution 
of the CAMELS galaxies are non-uniform when considering individual galaxies. 
We correct for this implicit prior by randomly selecting 100 galaxies from each 
simulation. 
This imposes a uniform prior on $\Omega$ and $\mathcal{B}$ and leaves us with 
a total of 100,000 CAMELS-TNG galaxies. 

Because our aim is to analyze the observed photometry of NSA galaxies, we forward
model photometry for the simulated galaxies. 
CAMELS-TNG provides synthetic unattenuated stellar photometry for each simulated galaxy. 
For each galaxy, the SED of every star particle in its host subhalo is modeled 
as a single-burst simple stellar population using stellar population synthesis 
(SPS) based on the recorded birth time, metallicity, and mass. 
The SPS uses FSPS~\citep{conroy2009}, Padova isochrones, MILES
stellar library, and assumes a \cite{chabrier2003} IMF. 
Then the SEDs of the star particles are combined to produce the unattenuated 
SED of the galaxy and convolved with SDSS $g$, $r$, $i$, $z$-band photometric
bandpasses to produce the unattenuated photometry. 

Next, we impose dust attenuation on the photometry. 
In the original TNG300 simulation, synthetic dust attenuated photometry was 
computed using a dust 
model based on the metal content of the neutral gas distribution in and around 
each galaxy~\citep{nelson2018}. 
We emulate this dust attenuation method in our forward model. 
First, we compile the unattenuated and attenuated SDSS $g$, $r$, $i$, $z$-band 
magnitudes for all TNG300 galaxies.   
We calculate the attenuation on the photometry by taking the 
difference between the unattenuated and attenuated magnitudes of 
the TNG300 galaxies. 
Then, for each CAMELS-TNG galaxy, we assign the attenuation of the TNG300 galaxy
with the closest unattenuated photometry based on L2 norm.
Each TNG300 galaxy has 12 sets of attenuated magnitudes, which correspond to 
the different observer viewing angles. 
We randomly select one of the viewing angles in the assignment.
We attenuate the CAMELS-TNG galaxies by the assigned attenuation to get the 
attenuated photometry: $X_i$. 

Finally, we add noise, $\sigma_X$, to the synthetic photometry based on the 
measured uncertainties of NSA galaxies. 
For each galaxy, we randomly sample $\sigma_{X,i}$ from the range of
uncertainties measured in NSA. 
Afterwards, we apply the uncertainty using a Gaussian with standard deviation
$\sigma_{X,i}$: $\hat{X}_i \sim \mathcal{N}(X_i, \sigma_{X, i})$. 
Although our noise model is simplistic, this is not an issue with our
approach. 
The posteriors we ultimately evaluate are conditioned on the measured 
uncertainties (right hand side of Eq.~\ref{eq:posterior}). 
Hence, similarly as in \cite{hahn2022a}, we only need to ensure that the
measured, $\hat{\sigma}_X$, of NSA galaxies is fully within the $\sigma_X$ 
distribution of our simulated dataset. 
In Figure~\ref{fig:nsa}, we present the color-magnitude distribution of the
forward modeled CAMELS-TNG galaxies (blue). 
The distribution is in good agreement with the distribution of NSA galaxies,
but slightly broader given the wide range of $\Omega$ and $\mathcal{B}$.
We reserve a random 10\% of these galaxies as a test dataset for validation
(Appendix~\ref{sec:valid}).

\begin{figure}[ht]
\begin{center}
    \centerline{\includegraphics[width=0.3\textwidth]{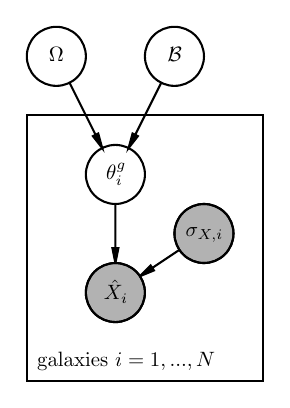}}
    \caption{
        Graphical representation of our hierarchical approach that illustrate
        the relationship among the parameters of our model. 
        White circles are inferred random variables and dark shaded circles are observed
        quantities.
        The physical properties of galaxies, $\theta^g_i$, are determined from
        the cosmological and hydrodynamical parameters $\Omega$ and
        $\mathcal{B}$ through CAMELS-TNG. 
        Then the noisy optical photometry, $\hat{X}_i$, is derived from
        $\theta^g_i$ using SPS and our noise model. 
    }\label{fig:graph}
\end{center}
\end{figure}

\section{Methods} \label{sec:methods} 
\subsection{Hierarchical Bayesian Population Inference} \label{sec:hier} 
Our goal is to infer the posterior of cosmological parameters
$\Omega = \{ \Omega_m, \sigma_8 \}$ and baryonic feedback parameters
$\mathcal{B}$ from the observed photometry of galaxies in the NSA catalog,
$\{\bfi X_i\}$:
$p(\Omega, \mathcal{B} \given \{{\bfi X_i}\})$.
${\bfi X_i}$ represents both the measured absolute magnitudes and
uncertainties: $\{ \hat{X}_i, \hat{\sigma}_{X,i}\}$. 
With our forward model we can simulate noisy galaxy photometry from $\Omega$
and $\mathcal{B}$. 
Hence, the cosmological inference from photometry can be reformulated as a
hierarchical population inference problem. 

To illustrate this, we graphically represent our forward model in
Figure~\ref{fig:graph}.
Circles and shaded circles represent random variables and observed quantities. 
$\theta_i^g$ represents the physical properties of galaxies (\emph{e.g.} $M_*$,
star-formation history), which are determined from $\Omega$ and $\mathcal{B}$
through CAMELS-TNG.
Then the noisy photometry $\hat{X}_i$ is determined from $\theta_i^g$ through
SPS and our noise model. 

We can rewrite the posterior as: 
\begin{align}
p(\Omega, \mathcal{B} \given \{{\bfi X_i}\}) 
    =&~\frac{p(\Omega, \mathcal{B})~p(\{{\bfi X_i}\} \given \Omega,
    \mathcal{B})}{p(\{{\bfi X_i}\})}\\
    =&~\frac{p(\Omega, \mathcal{B})}{p(\{\bfi X_i\})}\prod\limits_{i=1}^N 
    p({\bfi X_i}\given \Omega, \mathcal{B})\\
    =&~\frac{p(\Omega, \mathcal{B})}{p(\{\bfi X_i\})}\prod\limits_{i=1}^N 
    \frac{p(\bfi X_i)\,p(\Omega, \mathcal{B}\given{\bfi X_i})}{p(\Omega,
    \mathcal{B})}\\
    =&~\frac{1}{p(\Omega, \mathcal{B})^{N-1}}\prod\limits_{i=1}^N p(\Omega,
    \mathcal{B}\given {\bfi X_i})
\end{align} 
Here, we assume that galaxies are i.i.d.
Since we impose uniform priors, $p(\Omega, \mathcal{B}) \propto 1$ (Sec.~\ref{sec:sims}):
\begin{align}
p(\Omega, \mathcal{B} \given \{{\bfi X_i}\}) 
    \propto &~\prod\limits_{i=1}^N p(\Omega, \mathcal{B}\given {\bfi X_i}).
    \label{eq:posterior}
\end{align} 
Up to a constant, we can evaluate $p(\Omega, \mathcal{B} \given \{{\bfi X_i}\})$ if we can
accurately estimate $p(\Omega, \mathcal{B}\given {\bfi X_i})$, the posterior
for any single galaxy under uniform priors.
\vspace{4mm}

\subsection{Neural Density Estimation} \label{sec:anpe}
One way to accurately estimate $p(\Omega, \mathcal{B}\given {\bfi X_i})$ is
by applying neural density estimation (NDE) to the CAMELS-TNG, which provides a
training dataset of 100,000 parameter-photometry pairs: 
$\{(\Omega, \mathcal{B}, {\bfi X_i})\}$.
With NDE, we can use this data to train a neural network $q$ with parameters
$\phi$ to estimate 
$q_\phi(\Omega, \mathcal{B}\given {\bfi X_i}) \approx p(\Omega, \mathcal{B}\given {\bfi X_i})$.
This type of simulation-based inference using NDE has been applied to a broad
range of astronomical applications~\citep[\eg~][]{wong2020, dax2021, zhang2021,
hahn2022d}. 

In this work, our NDE is based on ``normalizing flow'' 
models~\citep{tabak2010, tabak2013, rezende2015}, which use neural 
networks to learn a flexible and bijective
transformation, $f$, that maps a complex target distribution to a simple base
distribution that is fast to evaluate.
$f$ is defined to be invertible and have a tractable Jacobian, so that the 
target distribution can be evaluated with change of variables. 
In our case, the target distribution is 
$p(\Omega, \mathcal{B}\given {\bfi X_i})$, and we set the base distribution to
be a multivariate Gaussian. 
Among different flow architectures, we use Masked Autoregressive
Flow~\citep[MAF;][]{papamakarios2017} models implemented in the $\mathtt{sbi}$
Python
package\footnote{\url{https://github.com/mackelab/sbi/}}~\citep{greenberg2019,
tejero-cantero2020}.

During training, we want to determine $q_\phi$ that best approximates 
$p(\Omega, \mathcal{B}\given {\bfi X_i})$. 
We reformulate this into an optimization problem of determining $\phi$
that minimizes the KL divergence between 
$p(\Omega, \mathcal{B}, {\bfi X_i}) = p(\Omega, \mathcal{B}\given {\bfi X_i})
 p({\bfi X_i})$ and
$q_\phi(\Omega, \mathcal{B}\given {\bfi X_i}) p({\bfi X_i})$.
In practice, we split the CAMELS-TNG data into a training and validation set
with a 90/10 split.  
Then, we maximize the total log-likelihood 
$\sum_i \log q_{\phi}(\Omega, \mathcal{B}\given {\bfi X_i})$ over the 
training set, which is equivalent to minimizing the KL divergence. 
To prevent overfitting, we evaluate the total log-likelihood on the validation
data at every training epoch and stop the training when the validation 
log-likelihood fails to increase after 20 epochs.  

We train a large number of flows ($\sim$2000) with architectures determined by the
\cite{akiba2019} hyperparameter optimization. 
We allow the numbers of blocks, transforms, and hidden neurons as well as the 
dropout probability and the learning rate to vary.
Then, we construct our final flow as an equally weighted ensemble of five
flows with the lowest validation losses: 
$q_{\phi}(\Omega, \mathcal{B}\given {\bfi X_i}) = 
\sum_{j=1}^5 q_{\phi,j}(\Omega, \mathcal{B}\given {\bfi X_i})/5$. 
Ensembling flows with different initializations and architectures improves the
overall robustness of our normalizing flow.
In Appendix~\ref{sec:valid}, we extensively validate the accuracy of $q_\phi$,
using Simulation-Based Calibration~\citep{talts2020} and the
\cite{lemos2023} coverage test. 
Our combined flow is a near optimal estimate of the true posterior. 

\begin{figure}[ht]
\begin{center}
    \centerline{\includegraphics[width=\textwidth]{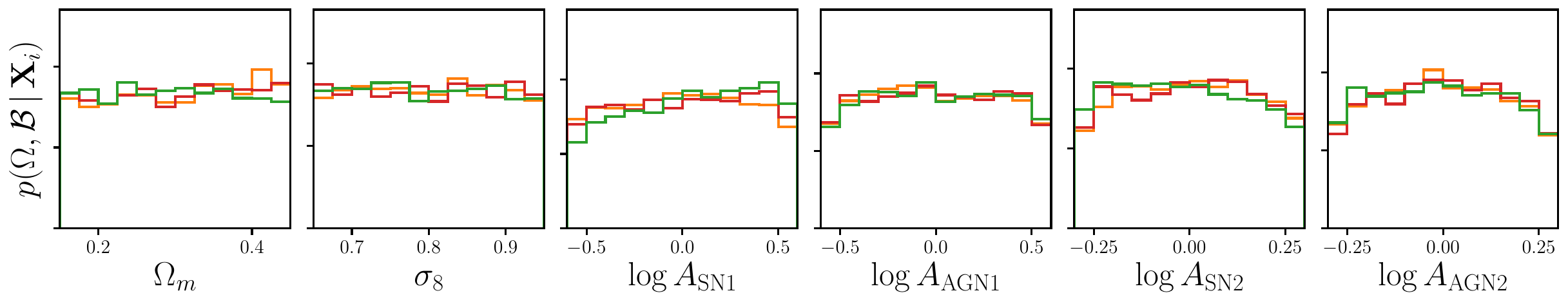}}
    \caption{Estimated marginalized posteriors of the cosmological and 
    hydrodynamical  parameters ($\Omega$, $\mathcal{B}$) from photometry, 
    $q_{\phi}(\Omega, \mathcal{B}\given {\bfi X_i})$, for three arbitrarily 
    selected NSA galaxies. 
    The posteriors correspond to the galaxies marked in Figure~\ref{fig:nsa}.
    The photometry of a single galaxy contains limited cosmological
    information. 
    }\label{fig:p_omega_x_i}
\end{center}
\end{figure}

In Figure~\ref{fig:p_omega_x_i}, we present $q_{\phi}(\Omega, \mathcal{B}\given {\bfi X_i})$
for three arbitrarily selected NSA galaxies. 
We mark the colors and magnitudes of the selected galaxies in Figure~\ref{fig:nsa} with 
crosses of the same color. 
The individual posteriors reveal that there is limited cosmological information in 
the photometry of a single galaxy. 
However, with Eq.~\ref{eq:posterior}
we can efficiently extract the cosmological  information from {\em  entire galaxy surveys}.
\newpage

\section{Results} \label{sec:results}
\begin{figure}[ht]
\begin{center}
    \centerline{\includegraphics[width=0.5\textwidth]{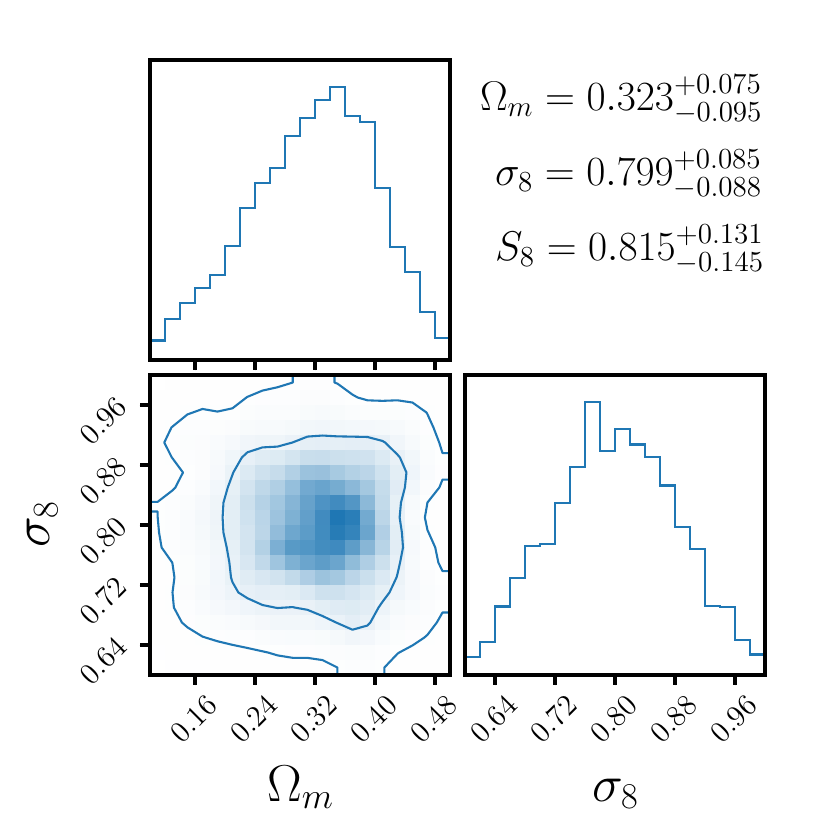}}
    \caption{
        The posterior of $\Omega_m$ and $\sigma_8$ inferred from the 
        observed photometry of 22,338 NSA galaxies.
        The contours mark the 68 and 95 percentiles. 
        We place significant cosmological constraints from the photometry of 
        galaxies alone. 
    }\label{fig:p_omega_x}
\end{center}
\end{figure}

With our trained NDE, $q_{\phi}(\Omega, \mathcal{B}\given {\bfi X_i})$, we can
now evaluate the posterior $p(\Omega, \mathcal{B}\given\{\bfi X_i\})$ for
multiple galaxies using Eq.~\ref{eq:posterior}. 
In Figure~\ref{fig:p_omega_x}, we present 
$p(\Omega, \mathcal{B}\given\{\bfi X_i\})$ for the 22,338 observed galaxies in
our NSA sample.
The contours mark the 68 and 95 percentiles of the distribution and we list the
median and $\pm1\sigma$ marginalized constraints on $\Omega_m$, $\sigma_8$, and
$S_8 = \sigma_8\sqrt{\Omega_m/0.3}$. 
The samples from $p(\Omega\given\{\bfi X_i\})$ are derived using Markov Chain
Monte Carlo (MCMC). 
We note that MCMC is necessary here since the individual posteriors 
$q_{\phi}(\Omega, \mathcal{B}\given {\bfi X_i})$ are combined multiplicatively 
(Eq.~\ref{eq:posterior}), we cannot separately sample 
$q_{\phi}(\Omega, \mathcal{B}\given {\bfi X_i})$ as this is equivalent to 
averaging the posteriors. 
We use the $\mathtt{emcee}$ sampler~\citep{foreman-mackey2013} with 1,000
walkers evaluated 100,000 iterations, discarding the first 10,000 iterations for
burn in.
This procedure requires ${\gg}10^5$ evaluations of the posterior {\em per galaxy}, 
which is computational prohibitive for standard sampling methods.  
However, NDE dramatically reduces this cost and make our approach computational
feasible.

Overall, we derive significant constraints on both $\Omega_m$ and $\sigma_8$: 
$\Omega_m = 0.323^{+0.075}_{-0.095}$ and $\sigma_8 = 0.799^{+0.088}_{-0.085}$. 
We also derive significant constraints on the hydrodynamical
parameters: 
$A_{\rm SN1} = 0.996^{+0.573}_{-0.377}$, 
$A_{\rm AGN1} = 0.987^{+0.603}_{-0.389}$, 
$A_{\rm SN2} = 1.003^{+0.340}_{-0.287}$, 
and 
$A_{\rm AGN2} = 1.017^{+0.396}_{-0.317}$. 
However, we do not include them in the figure for clarity.  
Our cosmological constraints demonstrate that although the photometry of a
single galaxy does not contain a significant amount of cosmological
information, we can place stringent cosmological constraints by combining the
information from as little as 22,338 galaxies. 

\section{Discussion} \label{sec:results}
A major caveat of our constraint is that our posterior assumes a galaxy
formation and a SED models. 
This assumption is more clearly expressed if we rewrite 
\begin{equation}
    p(\Omega, \mathcal{B}\given {\bfi X_i}) \propto p(\Omega, \mathcal{B}) \int     p({\bfi X_i} \given \theta^g_i)\, p( \theta^g_i\given\Omega, \mathcal{B})\,
    {\rm d}\theta^g_i.
\end{equation}
In our case, $p(\theta^g_i\given\Omega, \mathcal{B})$ is the TNG simulation 
and $p({\bfi X_i}\given\theta^g_i)$ is our forward model for photometry based 
on SED modeling. 

TNG is a state-of-the-art hydrodynamical simulation. 
Yet it makes specific choices and assumptions, \eg~for its subgrid treatments for
the formation and evolution of stellar populations, black hole growth, radiative 
cooling, stellar and black hole feedback, and magnetic 
fields~\citep{pillepich2018, nelson2018}.
Comparisons among hydrodynamical simulations \citep[\eg][]{hahn2019, dickey2021}, 
suggest that an alternative galaxy formation model may produce a significantly 
different $p(\theta^g_i\given\Omega, \mathcal{B})$. 

Similarly, the SED modeling, $p({\bfi X_i}\given\theta^g_i)$, includes a number of 
choices and assumptions. 
For instance, we use a dust model that assumes that dust is cospatial and
uniformly mixed. 
Yet, recent studies suggest that the dust-star geometry can be significantly
more complex and, thus, produce a wider range of attenuation
curves~\citep{narayanan2018a, hahn2021}. 
The SED model also does not accurately model nebular emission lines,
which may significantly impact the photometry of emission line galaxies. 
Furthermore, our SED model assumes a fixed Chabrier IMF. 
Observations, meanwhile, suggest that there may be significant 
variation (see \citealt{smith2020imf} for a recent review).

A key advantage of our approach is that we can systematically choose specific 
galaxy populations that are most robust to variations in galaxy formation and SED models. 
For example, we can select galaxies that have consistent cosmological information, 
as predicted by multiple galaxy formation models. 
As \cite{villaescusa-navarro2022} argue, the constraining power in galaxy properties 
on $\Omega_m$ comes from measuring the baryon 
fraction, $\Omega_b/\Omega_m$. 
This corresponds to measuring $M_{\rm tot}/M_b$ or $M_{\rm tot}/M_*/\epsilon_*$, 
where $\epsilon_*$ is the ``star forming efficiency''. 
Galaxy formation models typically calibrate $M_{\rm tot}/M_*$ against
constraints on the stellar-to-halo mass relation from 
observations~\citep[\eg][see also \citealt{wechsler2018} for a review]{mandelbaum2006, behroozi2010, moster2010, leauthaud2012, tinker2013, zu2015, gu2016}.
Hence, selecting galaxies that are robust to a change of the galaxy formation model
boils down to selecting galaxies with consistent $\epsilon_*$ across models. 

\begin{figure}[ht]
\begin{center}
    \centerline{\includegraphics[width=0.5\textwidth]{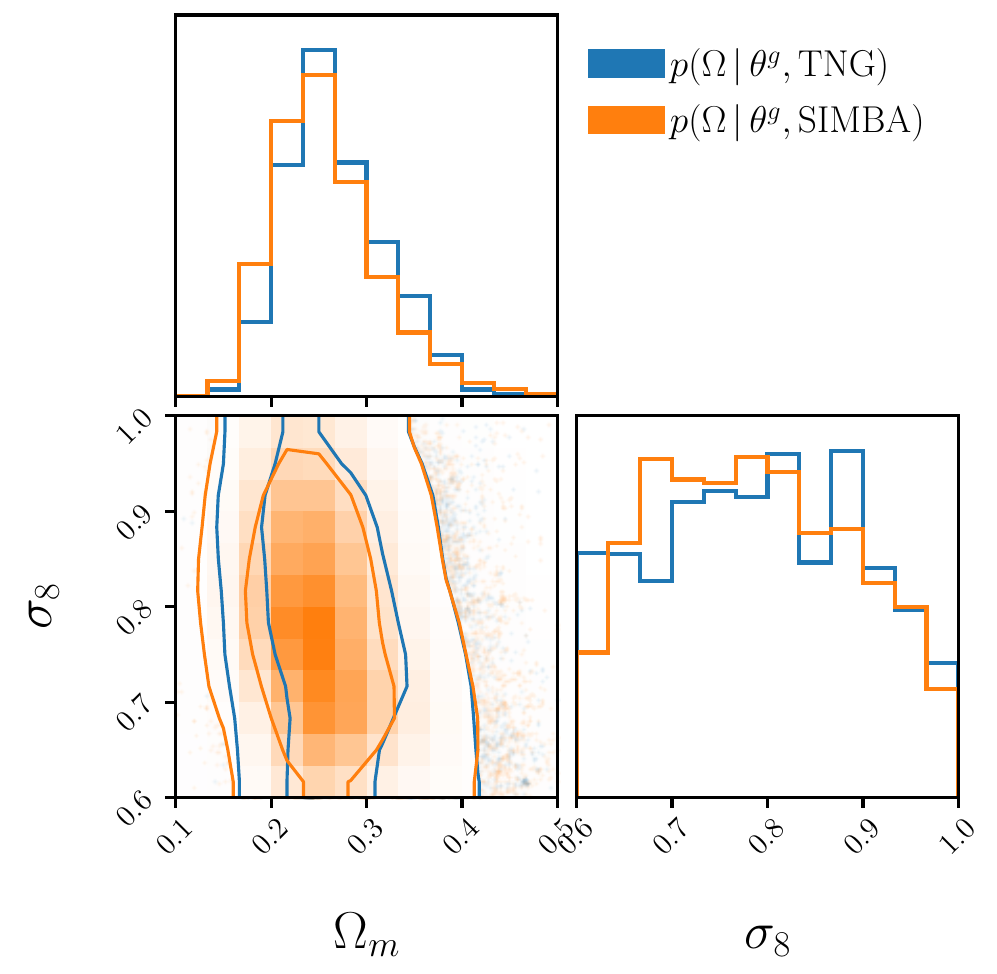}}
    \caption{
        The cosmological information, $p(\Omega\given\theta^g)$, of a star-forming 
        galaxy with $M_*\sim10^{9.5}M_\odot$ and ${\rm SFR} = 10^{-0.3}M_\odot/{\rm yr}$
        from TNG (blue) and from SIMBA (orange).
        The contours mark the 68 and 95 percentiles. 
        Specific galaxy populations, such as intermediate mass 
        star-forming galaxies 
        contain consistent cosmological information across different galaxy 
        formation models. 
    }\label{fig:p_omega_theta}
\end{center}
\end{figure}

One such population would be star-forming galaxies at intermediate $M_*$, whose 
relationship between gas and star formation are typically set by the
empirical \cite{kennicutt1989} relation~\citep[\emph{e.g.}][]{springel2003, dave2016}. 
To demonstrate this, in Figure~\ref{fig:p_omega_theta}, we compare the cosmological 
information content, $p(\Omega\given \theta^g_i)$, of a star-forming galaxy in TNG and 
a star-forming galaxy in SIMBA with similar galaxy properties: 
$M_*\sim 10^{9.5}M_\odot$ and 
${\rm SFR}\sim 10^{-0.3}M_\odot/{\rm yr}$. 
The TNG and SIMBA galaxies contain highly consistent cosmological information.

Another possible way to identify galaxies that are robust to a change of 
the galaxy formation model is to find galaxies that lie well within the support a 
given galaxy formation model. 
\cite{echeverri2023} showed that if you take galaxies from one galaxy 
formation model (\eg~TNG) and remove all the galaxies that are outliers 
of the model. 
The galaxies left have consistent cosmological information across multiple other 
galaxy formation models (\eg~SIMBA, Astrid, and Magneticum). 
\cite{echeverri2023} identified the outlier galaxies using their physical properties, 
but this approach could be applied to observables (photometry) as well. 

We can take a similar approach for the SED modeling assumptions by selecting
galaxies without emission lines and with limited dust attenuation, \eg~as probed 
by infrared photometry. 
We can also identify galaxy populations with observationally well constrained
IMFs or with little IMF variation~\citep{myers2013, smith2015a}.
Alternatively, we can also allow the parameters of the SED model to vary and 
incorporate them in our hierarchical inference. 
This would broaden our constraints overall; however, 
\cite{Cosmo1gal_SB28} promisingly find that the degradation in cosmological 
information of a single galaxy is limited even when allowing the 
IMF, and the vast majority of subgrid parameters in TNG, to vary. 

Our posterior in Figure~\ref{fig:p_omega_x} places significant 
constraints on $\sigma_8$. 
In contrast, previous studies find limited constraining power on $\sigma_8$ from 
the properties of a single galaxy~\citep{villaescusa-navarro2022, echeverri2023}. 
As Figure~\ref{fig:p_omega_theta} illustrates, however, the constraining power 
on $\sigma_8$ is non-negligible. 
Furthermore, our constraint is consistent with the fact that $\sigma_8$ constraints 
tighten significantly when using the properties of multiple 
galaxies~\citep{chawak2023_camels}.
Population statistics, such as the stellar mass or luminosity function, are 
sensitive to $\sigma_8$. 
With multiple galaxies, we can effectively exploit this constraining power.

Lastly, we note that Eq.~\ref{eq:posterior} does not include the selection
function, $S$, applied to our NSA sample. 
In principle, we can include $S$ in our formulation and infer 
$p(\Omega, \mathcal{B}\given \{\bfi{X}_i\}, S)$. 
Given our uniform prior, $p(\Omega,\mathcal{B}) \propto 1$, reformulating Eq.~\ref{eq:posterior} with $S$ reduces to 
including an additional factor of $p(\Omega, \mathcal{B}\given S)$. 
In this work, our selection is broadly set to remove any artifacts or objects
with problematic photometry and to ensure that the NSA sample is within the 
support of our simulated galaxies.
Including $p(\Omega, \mathcal{B}\given S)$ has a negligible impact on our
posterior. 

\section{Outlook} 
In this work, we selected $\sim$20,000 galaxies out of $\sim$120,000 in the 
NSA catalog, which provides photometry and spectroscopic redshifts.
Upcoming spectroscopic surveys such as the Dark Energy Spectroscopic
Survey~\citep[DESI;][]{abareshi2022} 
Bright Galaxy Survey~\citep[BGS;][]{hahn2022c} will observe 
{\em orders of magnitude} larger galaxy samples. 
With the tens of millions of galaxies that they will observe, these surveys will
create enormous sample sizes even when we restrict the galaxy populations to
those that are easy to model. 
This work can be directly extended to the observations from these surveys and 
leverage their unprecedented statistical power. 

We can also extend this work to upcoming photometric surveys, namely the Vera C. Rubin 
Observatory~\citep{ivezic2019}.
In this work, we relied on spectroscopic redshifts to limit our observation 
sample to a narrow redshift range. 
However, we can instead include $z$ as an additional parameter in $\theta^g$ and use 
forward modeled galaxies over a range of redshifts. 
\cite{li2023popsed} recently showed that the joint distribution of galaxy properties and 
redshift for a galaxy population can be simultaneously inferred from their photometry. 
This suggests that even with photometric redshifts, we can leverage the information on
galaxy properties to constrain cosmology. 
Furthermore, with photometric surveys, we would have access to the cosmological information 
in {\em hundreds of millions} of galaxies.

Beyond extending our work to larger samples, we can also include a wider range of observables.
In this work, we focus on optical broadband photometry. 
However, other observable encode additional information on the physical properties 
of galaxies, and thus cosmological information. 
For instance, emission lines measured from galaxy spectra can characterize the star 
formation history and gas content of galaxies.
In fact, even the continuum of galaxy spectra contain significant constraining power on 
galaxy properties over photometry~\citep{hahn2023_provabgs}. 
Beyond optical, radio observations of neutral hydrogen can also constrain
$V_{\rm max}$ from rotation curves~\citep[\eg][]{rogstad1971_hi, allen1973_hi, deblok2008_hi}.
Incorporating these observables would increase the precision of individual posteriors and, thus, 
has the potential to produce precise cosmological constraints with relatively smaller galaxy samples.
However, to leverage their constraining power, we would need a forward model that 
can accurately model them. 
So far, this remains a major obstacle.


\section*{Acknowledgements}
It's a pleasure to thank 
Shirley Ho, 
Pablo Lemos, 
Christopher C. Lovell, 
Benjamin Wandelt,
and Risa Wechsler for valuable discussions. 
This work was supported by the AI Accelerator program of the Schmidt Futures Foundation. 
This work was substantially performed using the Princeton Research Computing resources at Princeton University, which is a consortium of groups led by the Princeton Institute for Computational Science and Engineering (PICSciE) and Office of Information Technology’s Research Computing.

This research was conceived at the Kavli Institute for Theoretical Physics during the 
``Building a Physical Understanding of Galaxy Evolution with Data-driven Astronomy'' program and
was, thus, supported in part by the National Science Foundation under Grants No. NSF PHY-1748958 and PHY-2309135. The CAMELS project is supported by the Simons Foundation and the NSF grant AST 2108078.

This project made use of SDSS-III data. 
Funding for SDSSIII has been provided by the Alfred P. Sloan Foundation, the Participating Institutions, the National Science Foundation, and the U.S. Department of Energy Office of Science. 
The SDSS-III web site is \url{http://www.sdss3.org/}. 

SDSS-III is managed by the Astrophysical Research Consortium for the Participating Institutions of the SDSS-III Collaboration including the University of Arizona, the Brazilian Participation Group, Brookhaven National Laboratory, Carnegie Mellon University, University of Florida, the French Participation Group, the German Participation Group, Harvard University, the Instituto de Astrofisica de Canarias, the Michigan State/Notre Dame/JINA Participation Group, Johns Hopkins University, Lawrence Berkeley National Laboratory, Max Planck Institute for Astrophysics, Max Planck Institute for Extraterrestrial Physics, New Mexico State University, New York University, Ohio State University, Pennsylvania State University, University of Portsmouth, Princeton University, the Spanish Participation Group, University of Tokyo, University of Utah, Vanderbilt University, University of Virginia, University of Washington, and Yale University.

\appendix
\section{Validating the Neural Density Estimator} \label{sec:valid}
Our posterior (Eq.~\ref{eq:posterior}) requires an accurate estimate of the
individual posterior from NDE: 
$p(\Omega, \mathcal{B}\given {\bfi X_i}) \approx q_\phi(\Omega, \mathcal{B}\given {\bfi X_i})$
(Sec.~\ref{sec:anpe}). 
To validate $q_\phi$, we use 10\% of the CAMELS-TNG data reserved for testing
and two validation methods: (1) Simulation-Based Calibration (SBC) and (2)
the ``distance to random point'' (DRP) coverage test. 

\begin{figure}[ht]
\vskip 0.2in
\begin{center}
    \centerline{\includegraphics[width=0.9\columnwidth]{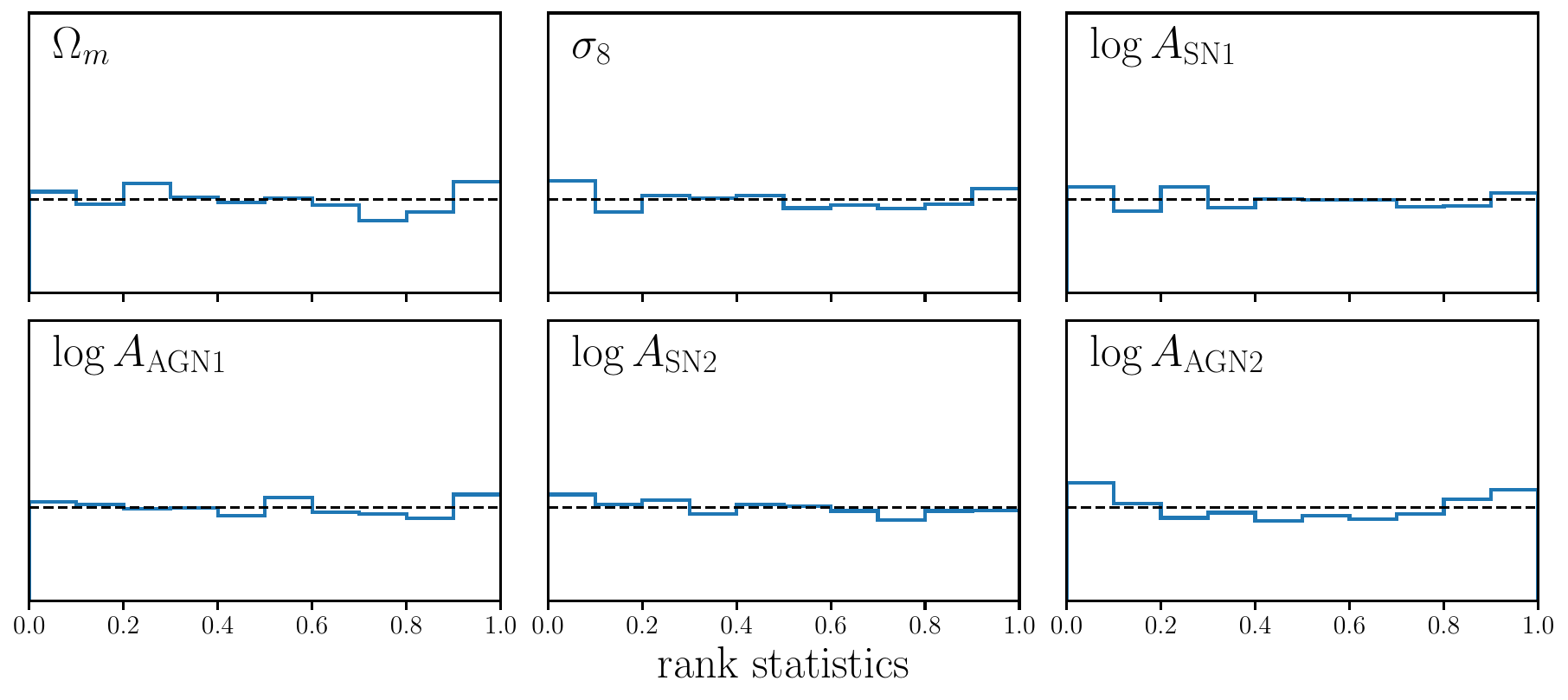}}
    \caption{
        Simulation-based calibration plot of 
        $q_\phi(\Omega, \mathcal{B}\given {\bfi X_i})$ using 10\% of the
        CAMELS-TNG data reserved for testing.
        The histogram in each panel represents the distribution of the rank
        statistic of the true value within the marginalized posterior (blue)
        for each parameter. 
        The rank distribution is uniform for the true posterior (black dashed).
        The rank distribution of $q_\phi$ is nearly uniform for all $\Omega$
        and $\mathcal{B}$ 
        parameters. 
        Therefore, it provides unbiased and accurate estimate of the true
        posterior.
    }\label{fig:ranks}
\end{center}
\vskip -0.2in
\end{figure}

Both are variations of the standard coverage test, where $q_\phi$ is applied to
test samples not used for training. 
The posterior of each test sample is compared against the true parameter value, 
then the percentile score of the true parameter is calculated.
Afterwards, cummulative distribution function (CDF) of the percentile is used
to assess the accuracy of $q_\phi$.
SBC is a modification of this standard coverage test that uses rank statistics
rather than percentile score. 
It addresses the limitation that the CDFs only asymptotically approach the true
values and that the discrete sampling of the posterior can cause artifacts in
the CDFs. 
In Fig.~\ref{fig:ranks}, we present the SBC rank distributions of $q_\phi$ for
$\Omega$ and $\mathcal{B}$ (blue). 
For the true posterior, rank distribution is uniform by construction (black
dashed).
The rank distributions are nearly uniform for all $\Omega$ and
$\mathcal{B}$.
Hence, we confirm that $q_\phi$ is in excellent agreement with the true
posterior.

\begin{figure}[ht]
\vskip 0.2in
\begin{center}
    \centerline{\includegraphics[width=0.4\columnwidth]{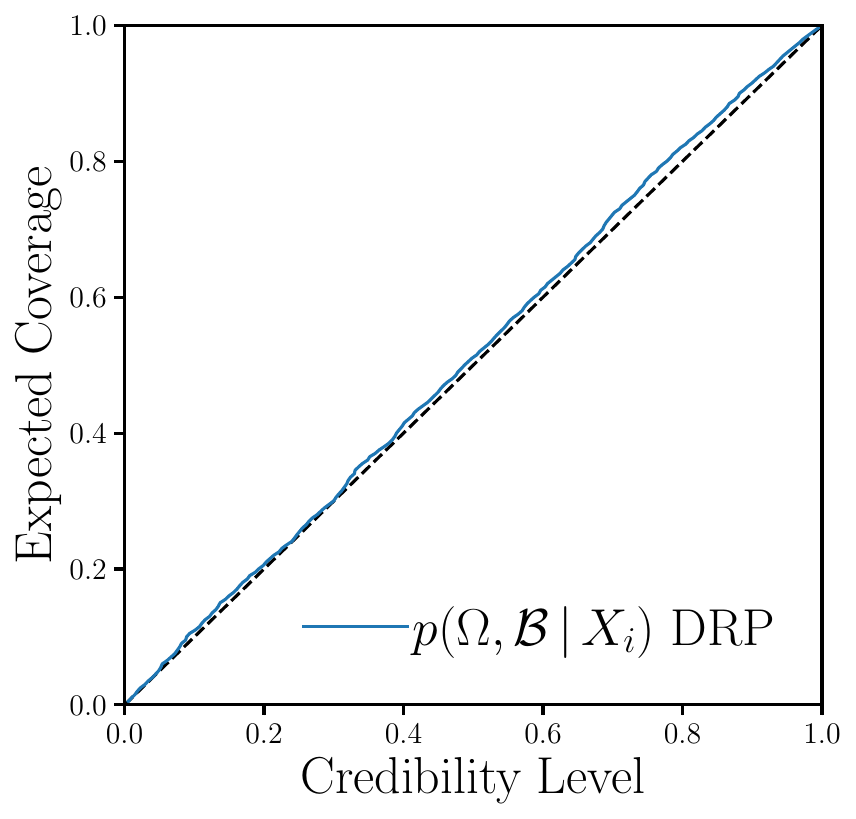}}
    \caption{DRP coverage test validating the accuracy of our 
    $q_{\phi}(\Omega, \mathcal{B}\given {\bfi X_i})$ posterior estimate (blue).
    The DRP test is calculated using  10\% of the CAMELS-TNG data reserved for
    testing.
    The black-dashed line represents an optimal estimate of the posterior.
    The DRP test demonstrates that $q_\phi$ provides a near optimal
    estimate of the true posterior.
    }\label{fig:tarp}
\end{center}
\vskip -0.2in
\end{figure}

As additional validation, we also use the DRP coverage test from
\cite{lemos2023}.
Instead of percentile scores or ranks, the DRP test assesses $q_\phi$ using
samples drawn from $q_\phi$ for a test sample, the true parameter value of the
test sample, and a random point in parameter space. 
It evalulates the distances between the $q_\phi$ samples and the random point. 
Then compares the distances to the distance between the true parameter value
and the random point in order to derive an estimate of expected coverage
probability. 
\cite{lemos2023} prove that this approach is necessary and sufficient to show
that a posterior estimator is optimal.
In Fig.~\ref{fig:tarp}, we present the DRP coverage test of $q_\phi$ (blue). 
Based on the DRP test, $q_\phi$ provides a near optimal estimate of the
true posterior (black-dashed). 

\begin{table*} 
    \centering
    \begin{tabular}{l|cc} 
        \hline
        color & 16\% & 84\% \\[3pt]
        \hline
        $g - r$ & 0.313 & 0.722 \\
        $g - i$ & 0.510 & 1.086 \\
        $g - z$ & 0.670 & 1.385 \\
        $r - i$ & 0.190 & 0.368 \\
        $r - z$ & 0.345 & 0.673 \\
        $i - z$ & 0.142 & 0.316 \\
        \hline            
    \end{tabular} \label{tab:color}
    \caption{The 16 and 84 percentiles of the color distributions for the forward
    modeled CAMELS-TNG galaxies. 
    We use these values as color cuts on the observed NSA sample (Section~\ref{sec:nsa}).
    This ensures that the observed sample lies within the support of our training data.} 
\end{table*}

\bibliographystyle{mnras}
\bibliography{goleta} 
\end{document}